**Commentary**

Manuscript title: **The changing dynamics of HIV/AIDS during the Covid-19 pandemic in the Rohingya refugee camps in Bangladesh – a call for action**

Muhammad Anwar Hossain[1,2], Iryna Zablotska-Manos[3, 4, 5]

1. Assistant Professor, Department of Sociology, Begum Rokeya University, Rangpur, Bangladesh.
2. College of Public Health Medicine and Veterinary Sciences, James Cook University, Townsville, Australia.
3. Westmead Clinical School, Faculty of Medicine and Health, University of Sydney, Westmead.
4. Marie Bashir Institute for Infectious Diseases and Biosecurity, University of Sydney, Westmead.
5. Western Sydney Sexual Health Centre, Western Sydney Local Health District, Parramatta, New South Wales, Australia.

**Corresponding author:** Muhammad Anwar Hossain
Assistant Professor, Department of Sociology, Begum Rokeya University, Rangpur, 5404, Bangladesh.
Phone: +8801815583933
E-mail: anwarsoc@brur.ac.bd

These authors have contributed equally to the work.

E-mail addresses of authors:
    MAH: anwarsoc@brur.ac.bd
    IZM: iryna.zablotska@sydney.edu.au

**Abstract**

**Introduction**

COVID-19 pandemic has affected each and every country's health service and plunged refugees into the most desperate conditions. The plight of Rohingya refugees is among the harshest. It has severely affected their existing HIV/STI prevention and management services and further increased the risk of violence and onward HIV transmission within the camps. In this commentary, we discuss the context and the changing dynamics of HIV/AIDS during COVID-19 among the Rohingya refugee community in Bangladesh.

**Discussion**

What we currently observe is the worst crisis in the Rohingya refugee camps thus far. Firstly, because of being displaced, Rohingya refugees have increased vulnerability to HIV, as well as to STIs and other poor health outcomes. Secondly, for the same reason, they have inadequate access to HIV testing treatment and care. Not only because of their refugee status but also because of the poor capacity of the host country to provide services. Thirdly, a host of complex economic, socio-cultural and behavioural factors exacerbate their dire situation with access to HIV testing, treatment and care. And finally, the advent of the COVID-19 pandemic has changed priorities in all societies, including the refugee camps. In the context of the unfolding COVID-19 crisis, more emphasis is placed on COVID-19 rather than other health issues, which exacerbates the dire situation with HIV detection, management, and prevention among Rohingya refugees.

**Conclusion**

Although the government of Bangladesh and different non-governmental organisations provide harm reduction, HIV care and COVID-19 care to refugees, a comprehensive response is needed to maintain and strengthen health programs for Rohingya refugees, including both HIV and COVID-19 care. This should include behavioural intervention, community mobilisation, and effective treatment and care. Without addressing the disadvantage of social conditions, it will be challenging to reduce the burden of HIV and COVID-19 among refugees. Despite the common crisis experienced by most countries around the world, the international community has an obligation to work together to improve the life, livelihood, and health of those who are most vulnerable. Rohingya refugees are among them.

# INTRODUCTION

In the context of a global crisis, it is the most vulnerable that suffer most. Refugees have limited access to health care in most of the host societies, including and especially access to HIV treatment, prevention, and care [1, 2]. COVID-19 pandemic has affected each and every country's health services and plunged refugees into the most desperate conditions. The plight of Rohingya refugees is among the harshest.

HIV/AIDS is one of the major global public health issues. Approximately 38 million people are living with HIV as of July 2021 [3]. It is one of the top five causes of mortality [4] and has taken around 33 million lives thus far [3]. It is estimated that 60 per cent of HIV cases are occurring among the key populations, including refugees [3].

To provide life-saving HIV/AIDS and sexual and reproductive health (SRH) services in humanitarian settings, the Inter-Agency Working Group (IAWG), in collaboration with the United Nations, non-governmental organisations and donors, developed the Minimum Initial Service Package (MISP) [5]. The focus of MISP objective 3 is to 'prevent the transmission of and reduce morbidity and mortality due to HIV and other STIs', including the identification of stakeholders for implementation, the prevention and management of sexual and gender-based violence, education aimed at prevention of HIV and STIs; counselling, prevention of newborn and maternal mortality and morbidity, prevention of unintended pregnancies and the delivery of comprehensive SRH services into primary health care [5]. However, the latest IAWG's (2012-2014) evaluation found that the quality and effectiveness of HIV/AIDS and other SRH services in humanitarian settings vary across different regions [6].

The Rohingya, who fled to escape state-sponsored genocide and violence in Myanmar, live in the world's largest and most densely populated refugee camps [7]. Bangladesh is home to around 1.2 million Rohingya [8]. Additionally, other countries also host Rohingya refugees, including Malaysia (over 102,00), Pakistan (55,000), India (over 40,000) and Nepal (600) [9, 10]. In refugee camps, Rohingya are facing serious challenges, including poor access to food, water, healthcare, sanitation, education, and livelihood [11, 12].

COVID-19 pandemic has disproportionately affected Rohingya refugees [12]. It has severely increased their vulnerability to poor health outcomes [13]. They are at high risk of contracting COVID-19 [14] because of the high density and poor sanitation within the camps [15, 16]. The rates of COVID-19 testing in the camps are very low, but the World Health Organization and other international agencies reported approximately 1354 confirmed cases and 18 deaths as of July 2021 in Bangladesh alone [17]. COVID-19 has severely affected the existing HIV/STI prevention and management services and further increased the risk of violence and onward HIV transmission within the camps.

Several studies have documented the health status, challenges, and wellbeing of Rohingya [18-22]. While some have examined such MISP areas as family planning [23], barriers and solutions for implementing maternal, newborn and child health programs [24], attitudes, experiences, coping strategies in relation to gender-based violence [25, 26], and comprehensive abortion care [27], there is limited evidence about the status of HIV/AIDS and other STIs among Rohingya [28]. In this commentary, we will discuss the context and the changing dynamics of HIV/AIDS during COVID-19 among the Rohingya refugee community in Bangladesh.

## DISCUSSION

### *HIV prevalence and care in Bangladesh: structural factors associated with poor access of refuges to HIV care*

In the general population of Bangladesh, the prevalence of HIV is very low at around 0.1 per cent [29]. However, there are approximately 14,000 people in Bangladesh living with HIV, including 1600 new cases in 2018 [30], and it is increasing among the high-risk groups such as people who inject drugs (IDU), men who have sex with men, male sex workers, transgender people, and male and female sex workers [31, 32]. It is reported that 38 per cent of new HIV cases are occurring among the IDUs due to syringe and needle sharing and unsafe sex [32].

Bangladesh is one of the poorest countries in the world and has limited capacity for HIV diagnosis and treatment. Treatment, support, and acre for the people living with HIV/AIDS are mainly provided by NGOs [31]. But they are struggling to provide continuous and uninterrupted services to patients due to the lack of resources and facilities in service delivery along with malnutrition and undernutrition of the HIV/AIDS patients. Only 50 per cent of identified HIV patients receive antiretroviral therapy (ART) [31]. The intervention program has reduced significantly, covering only 25.4 per cent of female sex workers and 23.6 per cent of MSM/MSW. Low coverage will create difficulties in ending HIV/AIDS by 2030 [31]. So, domestic capacity to provide HIV testing, treatment and care services in Bangladesh is limited. This situation has been further exacerbated by the unfolding COVID-19 pandemic, which has already resulted in at least 1,179,827 COVID-19 cases and 19,521 deaths in Bangladesh as of July 2021 [33].

*Rohingya refugees in Bangladesh: HIV prevalence and access to care*

Due to the large scale forced displacement, there is no systematic data about the health of Rohingya in host countries, as well as about their health in the home country where they were deprived of civic and human rights and citizenship [34]. Among Rohingya, it is assumed that around 5000 HIV infected people arrived in Bangladesh as the HIV prevalence rate in Myanmar was 0.8 per cent [35]. However, the levels of case detection are much lower: 273 cases were identified in 2018 and 319 in 2019, and only 277 people with HIV received treatment and care, while 19 died due to HIV/AIDS [35].

In Bangladesh, Rohingya's access to HIV/STI testing and treatment is severely limited. In the refugee camps, health services are provided by mainly international humanitarian agencies (including 62 international organisations and 8 UN bodies) and 59 national NGOs [36, 37]. These organisations operate 170 basic health units (which cover only one in 7,647 people in need), 33 primary health centres (for one in 39,394 people in need) and ten secondary care facilities (covering one in 130,000 people in need) [37]. Therefore, the indicators of HIV detection, treatment and care cascade among the Rohingya refugees in Bangladesh are among the poorest globally. Furthermore, in the context of the unfolding COVID-19 crisis, more emphasis is placed on COVID-19 rather than other health issues, which exacerbates the dire situation with HIV detection, management, and prevention among Rohingya refugees.

*Factors underlying poor access of Rohingya refugees to HIV care*

There are multiple interrelated economic, socio-cultural, and behavioural factors which are responsible for poor access of Rohingya refugees to HIV care.

Forced displacement has destroyed Rohingya refugee's livelihood, food security, income, and health [38, 39]. In the host country, they are forced to live in the high population density slums [15] with no sources of income. Approximately 66 per cent of households have no permanent sources of income; most live below the poverty line and are excluded from mainstream society [40]. By the end of 2020, as the result of the COVID-19 pandemic, their household income further decreased by around 20 per cent [41]. These deteriorating economic conditions further exacerbated the conditions for transmission of HIV/AIDS.

The majority of Rohingya refugees are illiterate, and very few can read and write. Awareness and knowledge about HIV and STIs are poor, and most Rohingya consider HIV an STIs' normal', like other diseases [42]. They rarely seek HIV/STI care, and condom use is generally low [43]. Cultural norms and conservative society prevent Rohingya from discussing sexual issues [45]. Sex and sexuality are considered taboo, a 'sin', while HIV and STIs - a 'curse of God', so humans cannot do anything about these diseases [37, 44, 45]. Levels of HIV/STI stigma and discrimination are high.

Gender analysis of the impact of HIV/AIDS on Rohingya refugees in Bangladesh is essential as women and children face a disproportionate burden in the new environment. UNHCR reported that 16 per cent of households are female-headed, 5 per cent of households are child-headed, and 4 per cent of households have people with disabilities [19, 46]. In a humanitarian setting, one out of every three displaced women and girls face sexual, physical and gender-based violence in their lifetime; one in five girls are married before 18 years old and have inadequate access to sexual and reproductive health services [47]. Child marriage and domestic violence are common; 74 per cent of Rohingya women had experienced gender-based violence, and 56.5 per cent - unwanted sexual intercourse with their husbands [48]. Gender inequality and gender-based violence increased during COVID-19 [49], which may also be indirectly responsible for the increased probability of HIV/STI transmission during the COVID-19 pandemic.

The international community and humanitarian agencies have already recognised the Rohingya as the most oppressed group in the world [50, 51]. While Rohingya fled genocide and extermination in their home country, they found themselves in a very vulnerable situation in the refugee camps in Bangladesh. Many have a severe traumatic experience and suffer stress and anxiety [48]. Some women experienced rape, gang rape, and the violent death of their family members [34]. Approximately 40,000 thousand pregnant women entered Bangladesh in 2017, a significant number of them being victims of rape [52], in most cases – gang rape by uniformed members of security forces [53]. There is also evidence of forced prostitution and trafficking [54]. Rohingya, especially those who live in the Cox's Bazaar district of Bangladesh, are well known to become victims of drug trafficking. Along with the host community, they are at high risk of drug use and HIV transmission [35]. Those who inject drugs are often excluded from mainstream society as the prevalence of stigma and discrimination against injecting drug use is high [32].

This generally desparate situation is further exacerbated by the COVID-19 pandemic. Poor knowledge about and fear of COVID-19 has been feeding stigma and discrimination against people with COVID-19, which has been seen as a curse of God [49].

**CONCLUSION**

What we currently observe is the worst crisis in the Rohingya refugee camps thus far. Firstly, because of being displaced, Rohingya refugees have increased vulnerability to HIV, as well as to STIs and other poor health outcomes. Secondly, for the same reason, they have inadequate access to HIV testing treatment and care. Not only because of their refugee status but also because of the poor capacity of the host country to provide services. Thirdly, a host of complex economic, socio-cultural, and behavioural factors exacerbate their dire situation with access to HIV testing, treatment, and care. And finally, the advent of the COVID-19 pandemic has changed priorities in all societies, including the refugee camps.

All of these factors not only raise Rohingya's vulnerability to HIV, but they can also facilitate an accelerated spread of HIV, and COVID-19, through the Rohingya refugee communities. The unfolding crisis associated with the COVID-19 pandemic is threatening to have a long-term effect on the livelihood and health of the most vulnerable populations, such as Rohingya refugees, including their vulnerability to HIV and poor access to HIV prevention and care.

Although the government of Bangladesh and different non-governmental organisations provide harm reduction, HIV care and COVID-19 care to refugees, a comprehensive response is needed to maintain and strengthen health programs for Rohingya refugees, including both HIV and COVID-19 care [51]. This should include behavioural intervention, community mobilisation, effective treatment and care for both infections, as well as COVID-19 immunisations. Without addressing the underlying dire social conditions, it will be challenging to reduce the burden of HIV and COVID-19 among refugees. Despite the common crisis experienced by most countries around the world, the international community has an obligation to work together to improve the life, livelihood, and health of those who are most vulnerable. Rohingya refugees are among them.

**Competing interests:** No potential conflict of interest was reported by the authors.

**Authors' contributions:** All authors contributed equally.

**Funding sources/sponsors:** Nil